# Electrical Characterization of DNA Origami Structures with PEG-PLL coating for improved robustness

Florian Heimbach[1,2,] Johann Bohlen[2], Hans Rabus[3], Woon Yong Baek[4,] Philip Tinnefeld[2]

## 1 Abstract

Electrical conductivity of DNA has been a controversial topic since it was first proposed in 1962. Disparities in the experimental results can often be explained by differences of the ambient or experimental conditions.

We report the dielectrophoretic trapping and electrical characterization of two rod-like DNA origami structures with different modifications. These were 12helix bundles, with either thiol-ends or polyA single-stranded overhangs, and 30-helix bundles, with thiol-ends and either with or without a PEG-PLL coating.

The observed impedance spectra showed ohmic resistances ranging from 500 kΩ to 3.4 MΩ for the 30-helix bundle structures with a PEG-PLL coating. DNA origami structures without the coating were always nonconductive. Depending on the structure, destruction of the gold nanoelectrodes occurred frequently during the trapping of DNA origami nanostructures. This indicates high currents, resulting from the trapped conductive nanostructures.

## 2 Introduction

Since the first proposed theory on DNA conductivity in 1962 [1], the question of whether or not DNA can be a conductor has been extensively investigated. As a mechanism for long-distance charge transport in DNA, a combination of π-orbital-overlapping and thermal hopping has been widely accepted [2]. Results of experimental studies still vary widely, ranging from insulating [3, 4, 5] to still conductive behavior [6, 7, 8, 9]. This, however, can often be explained by varying experimental conditions. The DNA to electrode interface and stability of the DNA helical structure are two major factors which need to be considered. Our goal is the fabrication of simple DNA circuitry for later application as radiation detector material [10]. The samples should have uniform impedance characteristics to allow comparable measurements. Dielectrophoretic trapping was chosen as preparation method. Repeatable, successful capture of single DNA molecules via dielectrophoresis (DEP) was reported by Kuzyk et al. [11]. Dielectrophoretic trapping would also allow the production of samples in adequate quantity.

The aforementioned problem of experimental conditions will be further examined in this paper. Thiol groups have been used successfully for binding DNA to gold electrodes with subsequent

[1] corresponding author; Physikalisch-Technische Bundesanstalt, Division 4 Optics, Bundesallee 100, 38116 Braunschweig
[2] LMU München, Department Chemie, Butenandtstr. 5 - 13, 81377 München
[3] Physikalisch-Technische Bundesanstalt, Division 8 Medical Physics and Metrological Information Technology, Abbestraße 2, 10587 Berlin
[4] Physikalisch-Technische Bundesanstalt, Division 6 Ionizing Radiation, Bundesallee 100, 38116 Braunschweig

measurable conductivity [12, 13, 14]. This might, however, still not be the optimal way as thiols are conventionally attached to the DNA backbone. Liu et al. [15] showed that thiolated nucleosides allow for even higher charge transport rates. These are, however, expensive to produce. A simpler way which has been tested is the employment of consecutive single-stranded adenine chains (polyA chains). These have been shown to adsorb onto gold surfaces with high affinity [16, 17, 18]. The polyA groups link the electrodes to the nucleoside's $\pi$-orbital and therefore potentially enable high charge transport rates.

DNA origami structures were chosen as sample material. We used a 12helix-bundle (12HB) in a honeycomb lattice (HCL) arrangement and a 30-helix bundle (30HB) in a square lattice (SQL) arrangement. Both have simple, rodlike shapes with 12 and 30 parallel DNA strands, respectively. Illustrations, as well as atomic force microscopy (AFM) and transmission electron microscopy (TEM) images of the DNA origami structures, are shown in figure 1. The three-dimensional structure of these DNA molecules offers two advantages. First, due to the tight packing of parallel strands, inner strands will be less influenced by ambient conditions such as low humidity and salt concentration. Second, it is expected that only the strands in direct contact with the surface should experience structural deformation.

Further, the 30HB was equipped with a protective polyethylene glycol (PEG)poly-L-lysine (PLL) coating. PLL coatings have been shown to preserve the origami structure in low salt concentrations and provides protection against nuclease degradation [19]. A similar protective effect of the layer might be expected from the effects of low solvation levels and DNA-surface interaction.

A previously observed phenomenon for dielectrophoretic trapping is the destruction of the nanometer-sized electrodes [14]. This hints towards conductive DNA structures: high electric currents can be identified as the cause of this destruction. When DNA origami structures were captured without electrode destruction, poor conductivity was observed [14]. This, however, is still in agreement with a theory of conducting DNA molecules. Most likely, these intact samples had compromised structures or were poorly contacted. Only low-conducting samples can yield nondestructive capture, since samples with higher conductivity would simply be destroyed.

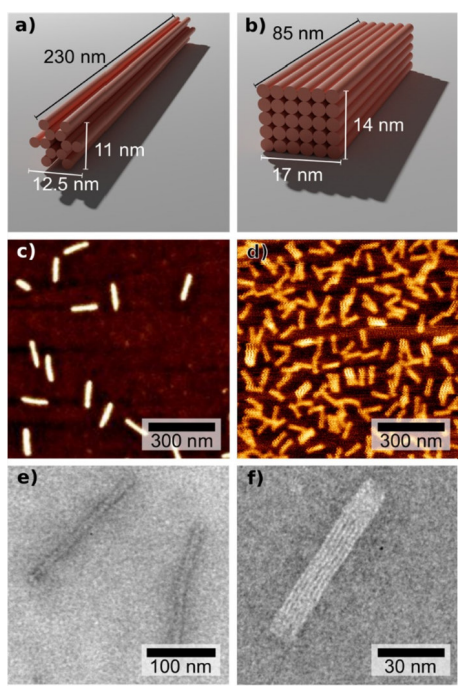

Figure 1: Illustrations of the 12HB and 30HB (a and b, respectively), as well as AFM (c and d) and TEM images (e and f) of DNA origami structures immobilized on surfaces.

| Parameter | 12HB PolyA | 12HB Thiol | 30HB Thiol | 30HB Thiol PEG-PLL |
|---|---|---|---|---|
| $U_{DEP}$ [$V_{rms}$] | 1 | 1 | 1 | 1 |
| $f_{DEP}$ [MHz] | 12.5 | 12.5 | 12.5 | 12.5 |
| $c_{DNA}$ [nM] | 5 | 5 | 1 | 1 |
| DEP duration [min] | 5 | 5 | 3 | 3 |
| Trapping fraction [% (#)] | 54 (41) | 43 (32) | 44 (35) | 17 (8) |
| Destruction fraction [% (#)] | 0 (0) | 11 (8) | 13 (10) | 54 (26) |

Table 1: Trapping parameters as well as trapping and destruction fractions for the capture via DEP of various DNA origami structures. The total number of experiments with successfully trapping, or where electrode destruction occurred, is listed in the brackets after the respective fraction. The DNA origami structure concentration $c_{DNA}$ was calculated assuming that no losses occurred during buffer exchange.

## 3 Results and Discussion

Trapping via DEP was observed for all investigated DNA origami sample material. The proportion of successful trials is called the trapping fraction and is listed in table 1. As expected, destruction of the electrodes occurred. The proportion of trials where this was the case is called the destruction fraction and is also listed in table 1.

The fractions varied depending on the DNA origami functionalization. The two thiolated, noncoated DNA origami samples had almost identical trapping and destruction fractions. Trapping with the polyA functionalized sample material did not yield any destroyed electrodes. The PEG-PLL-coated 30HBs, on the other hand, yielded the highest fraction of destroyed electrodes. Here, intact electrodes with captured origamis were obtained on average only for every sixth trial.

AFM images of DNA origami structures captured between two gold electrodes are shown in figure 2. The cross-sectional height profile along the blue arrows indicate capture of one DNA origami structure in figure 2a and of two parallel DNA origami structures in figure 2b. Figure 2c shows the remainders of a pair of electrodes that were eroded during trapping via DEP. In all cases shown, however, no electrical conductivity was observed. The impedance spectra were identical to that of an empty nanoelectrode chip (see figure 4).

Assuming that destroyed electrodes are a result of high conductivity in well contacted origami structures, we interpret these results as follows: PolyA chains did not provide suitable contacts for charge transport. A reason might be that the nucleotides closest to the DNA origami structure did not bind to the gold.

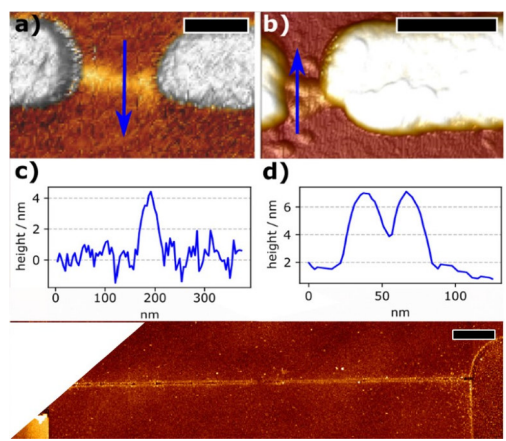

**Figure 2:** AFM images of captured DNA origami structures and destroyed fingertip electrodes. The underlying substrate in all images was silicon dioxide. Pictures **a** and **b** show successfully captured 12HB and 30HB, respectively, between two gold electrodes on a SiO$_2$ surface. The graphs in **c** and **d** show the height profiles along the blue arrow markers in **a** and **b**, respectively. **e** shows a pair of electrodes which were eroded during trapping via DEP. The scale bars (black rectangles) are 200 nm long in **a** and **b** and 1 m in **c**.

These would therefore be an obstacle for charge carriers. The thiol groups did provide ample contacts for charge transport. During trapping, the DNA origami structures are fully submerged in water. Therefore, it should not matter whether the DNA origami structure has an SQL or HCL arrangement. This explains the identical trapping fractions of the 12HB and uncoated 30HB samples. Still, the higher destruction fraction of the PEG-PLL-coated 30HB indicates that the uncoated samples might suffer from low salt denaturation. The interaction with the substrate surface might also be an influencing factor.

To avoid electrode destruction, a series resistor of 39 kΩ was used. Voltage losses were compensated to provide a 1 *V* peak-to-peak voltage at the electrode gap. In addition, "arrowhead" shaped electrodes were used. Due to their tapered shape, these electrodes have a higher cross-sectional area and therefore a higher ampacity than the fingertip electrodes. Sample preparation via DEP was performed with the 30HB and coated 30HB samples. Again, successful trapping, as well as destroyed electrodes, were observed, as seen in figure 3.

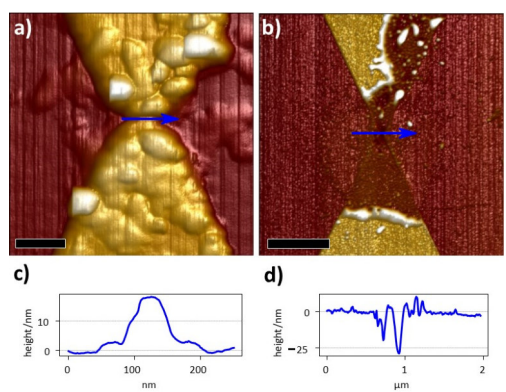

**Figure 3:** Arrowhead electrodes after trapping attempts via DEP with a 39 kΩ series resistor and coated 30HB structures. The underlying substrate in all images was silicon dioxide. Picture a shows a successful trapping and b a destroyed electrode. The scale bars (black rectangles) are 200 nm long in a and 2 m in b. The graphs in c and d show the height profiles across the blue arrow markers.

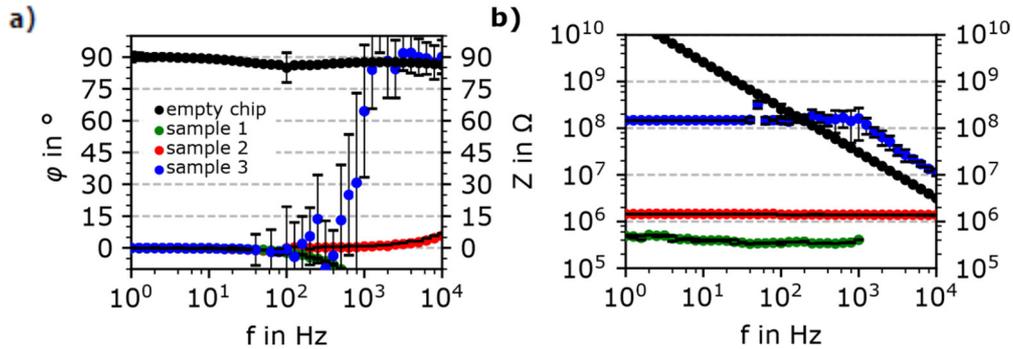

Figure 4: Impedance spectra of captured coated 30HB samples and an empty electrode chip. Displayed are the phase (a) and impedance (b) values for various chips with captured PEG-PLL-coated 30HB structures and an empty nanoelectrode chip (black dots) between 1 Hz and 10 kHz..

While all samples produced by trapping uncoated 30HB showed no signs of ohmic conductivity, measurable resistance values were registered in the case of PEG-PLL-coated DNA origami structures. Impedance spectra of some of these latter samples are shown in figure 4. The spectra show ohmic conductivity with resistance values as low as 500 kΩ, but typically in the MΩ range. The ohmic sample material forms a parallel circuit with the capacitive electrode chip. In such a circuit, the capacitive properties begin to take over for higher frequencies. This trend can already be seen to start at around 100 Hz, for the lowest conducting sample 4 with 150 MΩ resistance.

Some conducting samples were tested for their dielectric strength. Increases in ohmic resistance were indeed observed above voltages of 100 mV (see figure 5). Higher voltages resulted in even higher increases in resistance. These were irreversible and most likely produced by partial destruction of the sample.

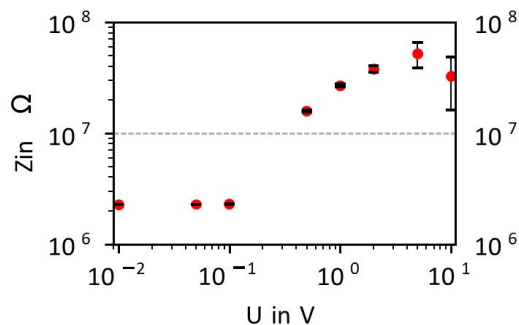

Figure 5: Dependence of the impedance values for a PEG-PLL-coated 30HB structure, captured between electrodes, on the applied voltage. The uncertainty bars represent the 95% confidence interval.

# 4 Conclusions

The trapping of thiolated DNA origami structures led to similar results as were reported by Shen et al. in 2015[14]. The thiolated, but uncoated DNA origami structures were electrically insulating in a dry environment. Shen et al. reported resistance values of several tens of 10 G$\Omega$ for uncoated, thiolized 30HB in a water-saturated environment. We observed no signs of electrical conductivity for these samples. Our conductivity measurements were, however, conducted in dry air. This observation further corroborates the claim that a sufficient humidity level is essential for DNA charge transport properties. In dry conditions, the helical B-DNA structure is less stable. This results in less regular pi-stacking and therefore decreasing charge transport rates.

Trapping of PEG-PLL-coated DNA origami structures resulted in different impedance spectra. We observed ohmic behavior in the M$\Omega$ range, four orders of magnitude lower than the uncoated structures reported by Shen et al. Apart from the fact that the electrodes were designed for higher ampacity, the key difference most likely arises due to the PEG-PLL coating of the 30HBs. One possible explanation would be that the current can flow through the PEG-PLL layer. L-lysine layers have been shown to be electrically insulating[20] and the same was expected for the PLL layer. PEG layers, on the other hand, can support an electric current[21, 22]. However, the conductivity values are several orders of magnitude too low, to explain the measured low resistance values.

A second possible explanation could be that the current can flow through the 30HBs. While all DNA origami structures may support high electrical currents in a suitable aqueous environment, dry conditions can alter that. Due to the PEG-PLL coating, the DNA origami structures are more robust under these dry conditions. In this context, a dense, multilayered structure, like the 30HB, is also expected to be advantageous. The stable DNA strands would support high charge transport rates. DNA origami structures can be prepared with PLL coating lacking the PEG layer[19]. While such a coating provides less protection of the DNA structure, the occurrence of ohmic conductivities in such structures would be evidence for the DNA based conductivity. Additionally, future experiments with coated 12HB structures could give further insight into the dominant charge transport mechanism.

Destroyed nanoelectrodes were still observed, even when using a current-limiting resistor. This further indicates towards the presence of highly conducting structures. A lower current limitation was not possible with the current setup, because a higher current limiting resistor would lead to higher voltage losses, which could no longer be compensated. In future experiments, it is planned to employ sapphire substrates. Electrode chips with this substrate material would form much weaker capacitors and therefore a highly increased range of resistor values can be employed. With this setup, it should be possible to trap even highly conducting structures via DEP without destroying the nanoelectrodes.

# 5 Materials and Methods

## 5.1 DNA Samples

Experiments were carried out with two different DNA origami structures: a 12helix bundle and a 30-helix bundle. Illustrations, AFM and TEM images of the structures, are shown in figure 1. A list of all samples can be found in table 1.

The short ends of all structures have functionalized groups for attachment to the gold electrodes. These linkage groups are either thiol or polyA functionalization. The thiolate sample have multiple oligonucleotides which are functionalized with up to two thiols. The thiols are connected via a C6 linker to the backbone of the DNA.

The sample material with polyA linkage groups possesses overhanging adenine chains. Each side of these DNA origami structures possessed twelve chains variing in length from six to 14 consecutive adenine bases. Additionally, the 30HBs were equipped with a PEG-PLL coating.

The 12HBs were produced at LMU Munich in the Tinnefeld lab [23]. The 12HBs were initially suspended in a storage buffer containing 40 mM Tris-base, 20 mM actic acid, 1 mM EDTA, and 20 mM $MgCl_2$.

The 30HBs were purchased from Tilibit nanosystems. The DNA origami structures were initially suspended in a storage buffer containing 5 mM Trisbase, 1 mM EDTA, 5 mM NaCl, and 20 mM $MgCl_2$.

## 5.2 AFM Imaging

AFM images were taken with a Bruker Dimension Icon in peak-force tapping mode. All measurements were done in dry air. Mica samples were prepared by first cleaving the mica using adhesive tape. The origami solutions were mixed with a 200 mM $MgCl_2$ buffer in a 1:1 ratio. A 10 l droplet was then suspended on the mica surface. After 5 min incubation time, the droplet was blown off in a nitrogen stream. The surface was then rinsed with ultrapure water and dried in a nitrogen stream.

## 5.3 Nanoelectrodes

In order to measure the electrical properties of molecules, they need to be connected to special electrodes. These bridge the microscopic molecules to our macroscopic measurement devices. Electrodes were fabricated by electron beam lithography, physical vapor deposition, and reactive ion etching [24].

Silicon wafers are used as a substrate material. These are (100) p-type, boron-doped wafers, with a resistivity of 1–10 $\Omega$cm and a total thickness of 381 m. The electrodes comprise a 20 nm gold layer, on top of a 5 nm platinum layer. Gold is used due to its poor chemical reactivity, while the platinum acts mostly as an adhesive layer. For insulation, a 600 nm silicon oxide layer is thermally grown on the silicon substrate. The fabrication was carried out at Physikalisch-Technische Bundesanstalt in Germany by the Weimann group.

The electrodes were designed in "fingertip" and "arrowhead" like shapes. Due to the randomness of the trapping process, special electrodes with multiple electrode pairs were used. These provide five electrode pairs in an electrically parallel arrangement. All electrode pairs were produced with either 120 nm or 80 nm gaps. The former were used for experiments with the 12HBs and the latter for experiments with the 30HBs. The impedance spectra and equivalent circuit of a nanoelectrode chip is shown figure 6.

## 5.4 Dielectrophoretic Trapping

The trapping is based on the AC dielectrophoretic effect. Successful capture via DEP of DNA molecules requires the buffer solution to be sufficiently nonconductive. This is due to electrothermal fluid flow [25] acting against the DEP forces [26]. Therefore, the DNA origami structures were transferred from their storage buffer into a trapping buffer solution. The buffer exchange was performed via spin filtering using Amicon Ultra centrifugal filters with a molecular weight cut-off of 100 kDa. 40 l origami solution and 460 l ultrapure water were pipetted into the filter cartridge. The cartridge was first spun with 14,000 G[1] for 3 min. Afterwards, the cartridge was refilled with ultrapure water and spun with 14,000 G for 3 min, refilled and spun again with 14,000 G for 5 min. The DNA origami solution was recovered by placing the filter upside down in a microcentrifuge tube and spun with 2,000 G for 5 min to extract the DNA origami structures from the filter. The final trapping buffer had a conductivity of 1.7 MΩcm.

The nanoelectrode chips were cleaned in an oxygen plasma prior to the trapping procedure. This removes light organic impurities from the surface. The electrodes were connected to the instrumentation via mechanical pressing with copper spring contacts. Trapping without a series resistor was carried out with a sinusoidal AC voltage $U_{DEP}$ applied to the electrodes. This voltage had a root-mean-squared (rms) value of 1 V [14] and a frequency $f_{DEP}$ of 12.5 MHz. The usage of a series resistor higher than 100 Ω requires higher applied voltages, because the capacitive characteristics of the chip cause a voltage loss over the series resistor. This can be compensated by applying correspondingly higher voltages. A 39 kΩ resistor was used with an applied rms voltage of 14.14 V and 12.5 MHz. The rms voltage at the electrode gap would therefore reach around 0.7 V. A 1 l droplet of DNA origami structures suspended in the trapping buffer was then applied on the nanoelectrodes. To prevent the droplet from drying during the DEP procedure, the sample stage was placed in a water-saturated environment. After a trapping time of 5 min, the voltage was shut off. The electrodes were short-circuited against each other. This prevents high-voltage stress due to electrostatic discharge while handling the sample stage. The sample was then dried in a nitrogen stream.

## 5.5 Impedance Spectroscopy

The measurement setup is designed for characterization of DNA via electrical impedance spectroscopy. DNA samples typically have high electrical impedance values. At the same time, the application of high voltages has to be avoided. This could lead to the destruction of the susceptible electrodes, or manipulation of the DNA itself [27]. These requirements can be met by using a lock-in amplifier. The AC voltage applied to the sample acts as the reference signal. The lock-in amplifier measures the current value at the same frequency.

For the measurements, a Zurich Instruments MFLI lock-in amplifier [28] was used. This device incorporates a lock-in amplifier as well as a signal generator. The reference signal for the lock-in amplifier is therefore internally provided. During the recording of an impedance spectrum, ten current measurements are performed per frequency value. A Gaussian fit is used to calculate the mean value and standard deviation for the current at the respective frequency. To avoid damaging the DNA molecules, the applied voltage was limited to 1 mV. This voltage was applied to the sample and the lock-in amplifier in series. The sample was mounted in a sample holder inside a metal housing. It was contacted via copper spring contacts. The impedance spectrum of the sample holder resembled that of a 3.5 fF capacitor. An unused, clean nanoelectrode chip has an impedance spectrum as shown in figure 6a and b. Such a spectrum can be reproduced using an equivalent circuit as shown in figure 6c. The simulated impedance spectrum was obtained using the SPICE software LTSpice from Linear Technology. Simulations were performed for a set of parameter values and the resistance and capacitance values given in figure 6c were those giving the best fit with the measured spectrum.

---

[1] 1 G = 9.81 $\frac{m}{s^2}$

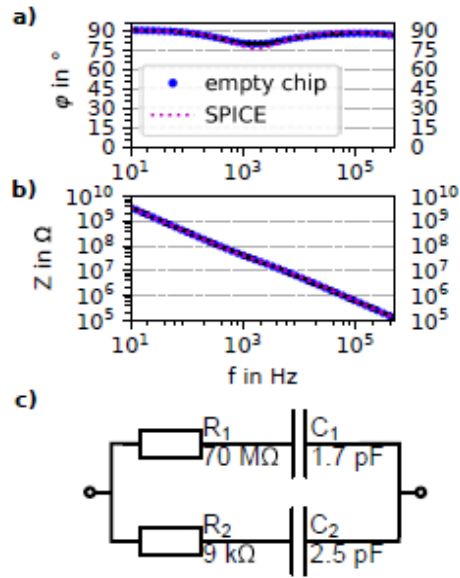

Figure 6: Impedance spectrum of an empty nanogap chip. Phase and absolute impedance data between 10 Hz and 500 kHz are shown in a and b, respectively. Measured data are shown as blue dots. A SPICE simulated equivalent circuit is shown in c. The results of the simulation are represented by the magenta dotted line.

# 6  Acknowledgment

We would like to express our gratitude to Heike Nittmann for her assistance in performing the sample preparation and measurements. We thank Alexander Ruhz from department 5.5 Scientific Instrumentation of PTB for his advice and help in preparing the samples and Frank Pohlenz from the department 5.2 Dimensional Nanometrology for supporting us with AFM measurements. We would also like to thank Dr. Jean-Philippe Sobczak from Tilibit Nanosystems for his advice regarding DNA origami structures. We gratefully thank Jussi Toppari from the University of Jyväskylä and Kosti Tapio from the University of Potsdam for supporting us with expertise regarding trapping via DEP.

# References

[1] D. D. Eley, G. D. Parfitt, M. J. Perry, and D. H. Taysum. The Semiconductivity Of Organic Substances. Part 1. *Transactions of the Faraday Society*, 49(0):79–86, 1953.

[2] Elizabeth M. Boon and Jacqueline K. Barton. Charge transport in DNA. *Current Opinion in Structural Biology*, 12(3):320–329, June 2002.

[3] Erez Braun, Yoav Eichen, Uri Sivan, and Gdalyahu Ben-Yoseph. DNAtemplated assembly and electrode attachment of a conducting silver wire. *Nature*, 391(6669):775–778, 1998.

[4] Y. Zhang, R. H. Austin, J. Kraeft, E. C. Cox, and N. P. Ong. Insulating Behavior of $\lambda$-DNA on the Micron Scale. *Physical Review Letters*, 89:198102, Oct 2002.

[5] Sungmin Hong, Luis A. Jauregui, Norma L. Rangel, Huan Cao, B. Scott Day, Michael L. Norton, Alexander S. Sinitskii, and Jorge M. Seminario. Impedance measurements on a DNA junction. *Journal of Chemical Physics*, 128(20):201103, May 2008.

[6] Hans-Werner Fink and Chrtian Schoeneberger. Electrical conduction through DNA molecules. *Nature*, 398:407–410, April 1999.

[7] Danny Porath, Alexey Bezryadin, Simon de Vries, and Cees Dekker. Direct measurement of electrical transport through DNA molecules. *letters to nature*, 403:635–638, February 2000.

[8] Hezy Cohen, Claude Nogues, Ron Naaman, and Danny Porath. Direct measurement of electrical transport through single DNA molecules of complex sequence. *Proceedings of the National Academy of Sciences*, 102(33):11589–11593, 2005.

[9] S. Kassegne, D. Wibowo, J. Chi, V. Ramesh, A. Narenji, A. Khosla, and J. Mokili. AC electrical characterisation and insight to charge transfer mechanisms in DNA molecular wires through temperature and UV effects. *IET Nanobiotechnology*, 9(3):153–163, 2015.

[10] Florian Heimbach, Alexander Arndt, Heidi Nettelbeck, Frank Langner, Ulrich Giesen, Hans Rabus, Stefan Sellner, J. Toppari, Boxuan Shen, and Woon Yong Baek. Measurement of changes in impedance of DNA nanowires due to radiation induced structural damage: A novel approach for a DNAbased radiosensitive device. *The European Physical Journal D*, 71:211, 08 2017.

[11] Anton Kuzyk, Bernhard Jurke, Jussi Toppari, Veikko Linko, and Paivi Torma. Dielectrophoretic trapping of DNA origami. *Small*, 4(4):447–450, April 2008.

[12] N. Kang, A. Erbe, and E. Scheer. Electrical characterization of DNA in mechanically controlled break-junctions. *New Journal of Physics*, 10(2):023030, feb 2008.

[13] Huijuan Zhang, Wei Xu, Xiaogang Liu, Francesco Stellacci, and John T. L. Thong. Capturing a DNA duplex under near-physiological conditions. *Applied Physics Letters*, 97(16):163702, 2010.

[14] Boxuan Shen, Veikko Linko, Hendrik Dietz, and J. Jussi Toppari. Dielectrophoretic trapping of multilayer DNA origami nanostructures and DNA origami-induced local destruction of silicon dioxide. *Electrophoresis*, 36:255–262, Jan 2015.

[15] S. P. Liu, J. Artois, D. Schmid, M. Wieser, B. Bornemann, S. Weisbrod, A. Marx, E. Scheer, and A. Erbe. Electronic transport through short dsDNA measured with mechanically controlled break junctions: New thiolgold binding protocol improves conductance. *physica status solidi (b)*, 250(11):2342–2348, 2013.

[16] Aric Opdahl, Dmitri Y. Petrovykh, Hiromi Kimura-Suda, Michael J. Tarlov, and Lloyd J. Whitman. Independent control of grafting density and conformation of single-stranded DNA brushes. *Proceedings of the National Academy of Sciences of the United States of America*, 104:9–14, Jan 2007.

[17] Sarah M. Schreiner, David F. Shudy, Anna L. Hatch, Aric Opdahl, Lloyd J. Whitman, and Dmitri Y. Petrovykh. Controlled and Efficient Hybridization Achieved with DNA Probes Immobilized Solely through Preferential DNASubstrate Interactions. *Analytical Chemistry*, 82(7):2803–2810, April 2010.

[18] Sarah M. Schreiner, Anna L. Hatch, David F. Shudy, David R. Howard, Caitlin Howell, Jianli Zhao, Patrick Koelsch, Michael Zharnikov, Dmitri Y. Petrovykh, and Aric Opdahl. Impact of DNA-Surface Interactions on the Stability of DNA Hybrids. *Analytical Chemistry*, 83(11):4288–4295, June 2011.

[19] Nandhini Ponnuswamy, Maartje M. C. Bastings, Bhavik Nathwani, Ju Hee Ryu, Leo Y. T. Chou, Mathias Vinther, Weiwei Aileen Li, Frances M. Anastassacos, David J. Mooney, and William M. Shih. Oligolysine-based coating protects DNA nanostructures from low-salt denaturation and nuclease degradation. *Nature communications*, 8:15654, May 2017.

[20] Zehra Durmus, Hüseyin Kavas, Muhammet Sadaka Toprak, Abdülhadi Baykal, Tuba Gürkaynak Altın¸cekic¸, Ay¸se Aslan, Ayhan Bozkurt, and Sedat Co¸sgun. l-lysine coated iron oxide nanoparticles: Synthesis, structural and conductivity characterization. *Journal of Alloys and Compounds*, 484(1):371–376, 2009.

[21] M. C. Wintersgill, J. J. Fontanella, P. E. Stallworth, S. A. Newman, S. H. Chung, and S. G. Greenbaum. Electrical conductivity, DSC and NMR studies of PEG and PPG containing lithium salts. *Solid State Ionics*, 135(1):155–161, 2000. Proceedings of the 12th International Conference on Solid State.

[22] O. Erdamar, Y. Skarlatos, G. Aktas, and M. N. Inci. An Experimental Work On The Electrical Conductivity Of PEG Under Changing Relative Humidity. *AIP Conference Proceedings*, 899(1):441–442, 2007.

[23] Jürgen J. Schmied, Mario Raab, Carsten Forthmann, Enrico Pibiri, Bettina Wünsch, Thorben Dammeyer, and Philip Tinnefeld. DNA origamibased standards for quantitative fluorescence microscopy. *Nature Protocols*, 9(6):1367–1391, June 2014.

[24] Th. Weimann, H. Scherer, H. Wolf, V. A. Krupenin, and J. Niemeyer. A new technology for metallic multilayer single electron tunneling devices. *Microelectronic Engineering*, 41-42:559–562, 1998. International Conference on Micro- and Nanofarbication.

[25] A. Castellanos, A. Ramos, A. González, N. G. Green, and H. Morgan. Electrohydrodynamics and dielectrophoresis in microsystems: scaling laws. *Journal of Physics D: Applied Physics*, 36(20):2584–2597, oct 2003.

[26] Kimmo Laitinen. Fluid dynamics in DEP trapping of DNA origamis and their functionalization. Master's thesis, University of Jväskylä, April 2009.

[27] Gianaurelio Cuniberti, Luis Craco, Danny Porath, and Cees Dekker. Backbone-induced semiconducting behavior in short DNA wires. *Physical Review B*, 65:241314, Jun 2002.

[28] Zurich Instruments AG, Zurich, Switzerland. *MFLI 500 kHz / 5 MHz Lock-in Amplifier*.

# 7   Graphical Abstract

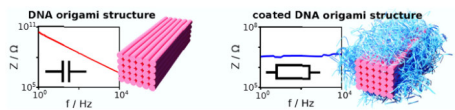

Mulitlayered DNA origami structures are speculated to preserve the regular helical structure of the DNA allowing for effective charge transport. We report the dielectrophoretic trapping of multilayered DNA origami structures between gold nanoelectrodes. Ohmic conductivity was measured in structures with PEGPLL coating whereas uncoated structures were electrically insulating.